\documentclass[aps,preprint,floats]{revtex4}
\usepackage{graphicx}
\begin{document}

\preprint{hep-ph/0404229}

\title{$K^+ \rightarrow \pi^+ \nu \bar{\nu}$ and
FCNC from non-universal $Z^\prime$ bosons}

\author{Xiao-Gang He}
\email{hexg@phys.ntu.edu.tw}
 \altaffiliation[On leave of absence from
]{Department of Physics, National Taiwan University, Taipei.}
\affiliation{%
Department of Physics, Peking University, Beijing}

\author{G. Valencia}
\email[]{valencia@iastate.edu} \affiliation{Department of Physics,
Iowa State University, Ames, IA 50011}

\date{\today}

\begin{abstract}

Motivated by the E787 and E949 result for 
$K^+ \rightarrow \pi^+ \nu \bar{\nu}$ we
examine the effects of a new non-universal right-handed
$Z^\prime$ boson on flavor changing processes. We place bounds on the
tree-level FCNC from $K-\bar{K}$ and $B-\bar{B}$ mixing
as well as from the observed CP violation in kaon decay. 
We discuss the implications for $K \rightarrow \pi \nu \bar{\nu}$, 
$B \rightarrow X \nu \bar{\nu}$ and $B\rightarrow \tau^+ \tau^-$. 
We find that the existing bounds allow substantial enhancements 
in the $K^+ \rightarrow \pi^+ \nu \bar{\nu}$ rate, particularly 
through a new one-loop $Z^\prime$ penguin operator.

\end{abstract}


\maketitle

\section{Introduction}

The experiments E787 and E949 at Brookhaven National Laboratory
have detected three events for the rare mode $K^+\rightarrow \pi^+
\nu \bar\nu$. Although the statistics are rather limited so far,
the collaborations have published a branching ratio
$B(K^+\rightarrow \pi^+ \nu \bar\nu) =
(1.57^{+1.75}_{-0.82})\times 10^{-10}$ for
E787~\cite{Adler:2001xv} and $B(K^+\rightarrow \pi^+ \nu \bar\nu)
= (1.47^{+1.30}_{-0.89})\times 10^{-10}$ for the combined E787 and
E949 \cite{Artamonov:2004hr} which is consistent within errors
with the standard model prediction $B(K^+\rightarrow \pi^+ \nu
\bar\nu) = (7.2\pm 2.1) \times 10^{-11}$
\cite{Buchalla:1995vs,smz,Battaglia:2003in} but which has a
central value roughly twice the standard model expectation.

This potential discrepancy with the SM has already generated much
interest in the literature. In extensions of the SM there are new
sources for flavor changing neutral currents which can affect the
branching ratio for $K^+\to \pi^+ \nu\bar \nu$ and reproduce the
central value obtained by E787 and E949~\cite{smz,susy,z,other}.

New interactions that affect predominantly the third generation are
a potential type of new physics that can also enter in this
context. Since the process occurs in the standard model via an
intermediate top-quark loop, and since it is sensitive to all
neutrino generations, $K \rightarrow \pi \nu \bar\nu$ has the
potential to probe for non-standard couplings of the third 
generation.

Several scenarios for interactions that single out the third
generation have been discussed in the literature. Among them
topcolor, where the low energy processes mediated by FCNC have
already been discussed in Ref.~\cite{Buchalla:1995dp}. Another
possibility consists of new gauge interactions for the third
generation. One concrete example was developed in
Ref.~\cite{He:2002kn,He:2003qv} where we constructed a class of
models with enhanced right-handed interactions for the third
generation. These models are motivated by the small disagreement
between the standard model and the measured $A_{FB}^b$ at LEP
\cite{Chanowitz:1999jj,Chanowitz:2001bv}. In this paper we will
use this example to illustrate how rare modes such as $K
\rightarrow \pi \nu \bar\nu$ indirectly probe the physics of the
third generation. We consider all the low energy FCNC processes 
that involve third generation leptons. Some of these processes  
constrain our models and we use them to discuss the implications 
for the rest.

The models we consider have been described in detail in
Ref.~\cite{He:2003qv}. Their distinctive characteristic is a
new right-handed interaction that affects predominantly the
third family. Models of this type are interesting in general 
because they provide specific parameterizations for potential
new physics in the couplings of the top-quark. These, in turn, 
are necessary to test the standard model with experiments 
designed to determine precisely how the top-quark
interacts. The specific form of the interactions in the models we
consider here is motivated by the apparent anomaly in the
measurement of $A^b_{FB}$ at LEP. 

\section{The Model}

The models discussed are variations of left-right models
in which the right-handed interactions single
out the third generation. The first two generations are chosen to
have the same transformation properties as in the standard model
with $U(1)_Y$ replaced by $U(1)_{B-L}$,
\begin{eqnarray}
&&Q_L = (3,2,1)(1/3),\;\;\;\;U_R = (3,1,1)(4/3),\;\;\;\;D_R =
(3,1,1)(-2/3),
\nonumber\\
&&L_L = (1,2,1)(-1),\;\;\;\;E_R = (1,1,1)(-2). \label{gens12}
\end{eqnarray}

The numbers in the first parenthesis are the $SU(3)$, $SU(2)_L$
and $SU(2)_R$ group representations respectively, and the number
in the second parenthesis is the $U(1)$ charge. 
For the first two generations this is the same as the 
$U(1)_Y$ charge in the SM and for the third generation 
it is the usual $U(1)_{B-L}$ charge of LR models.

The third generation is chosen to transform differently,
\begin{eqnarray}
&&Q_L(3) = (3,2,1)(1/3),\;\;\;\;Q_R(3) = (3,1,2)(1/3),\nonumber\\
&&L_L(3) = (1,2,1)(-1),\;\;\;\;L_R = (1,1,2)(-1). \label{gen3}
\end{eqnarray}

The correct symmetry breaking and mass generation of particles can
be induced by the vacuum expectation values of three Higgs
representations: $H_R = (1,1,2)(-1)$, whose non-zero vacuum
expectation value (vev) $v_R$ breaks the group down to
$SU(3)\times SU(2)\times U(1)$; and the two Higgs multiplets, $H_L
= (1,2,1)(-1)$ and $\phi = (1,2,2)(0)$, which break the symmetry
to $SU(3)\times U(1)_{em}$.

The relative strength of left- and right-handed interactions
is determined by a parameter $\cot\theta_R$. In the 
limit in which this parameter is large, the new right-handed
interactions affect predominantly the third generation. The
right-handed interactions may have a triplet of gauge bosons or
only a $Z^\prime$ depending on the model. The $W^\prime$ mixes with the 
$W$ through a mixing angle $\xi_W$ which is severely constrained 
by $b\rightarrow s \gamma$ \cite{He:2002kn}. The $Z^\prime$ will also mix
with the $Z$ through a mixing angle $\xi_Z$, and this mixing is 
severely constrained as well. In particular, it was found in
Ref.~\cite{He:2003qv} that the measurement of $g_{R\tau}$ at
LEP\cite{Abbaneo:2001ix} implies
\begin{equation}
\cot\theta_R \xi_Z \leq 10^{-3}
\label{xizcon}
\end{equation}
if the new interaction affects the third generation leptons as well
as the quarks. One way out of this problem is to extend the models with
additional fermions so that the interactions of the $\tau$ and 
$\nu_\tau$ are the same as in the standard model. These extensions are not
significantly constrained by $K\rightarrow \pi \nu \bar{\nu}$ or 
other low energy FCNC processes 
and we will not discuss them further in this paper.

Alternatively one must demand that the $W-W^\prime$ and $Z-Z^\prime$ 
mixing angles be small. A simple way to accomplish this at tree-level 
has been discussed in Ref~\cite{He:2003qv}. It consists of taking the vevs 
of the $\phi$ field to be $v_2=0$ in the $2-2$ entry 
and $v_1=v$ in the $1-1$ entry. This eliminates $W-W^\prime$ 
mixing. In addition, one can also eliminate $Z-Z^\prime$ mixing 
at tree level by demanding that the vev of $H_L$, $v_L$, be
equal to $v \cot\theta_R$. 

In Ref.~\cite{He:2003qv} the process $e^+e^- \rightarrow b \bar{b}$ at
LEP-II was used to obtain a lower bound for the mass of the new
$Z^\prime$ gauge boson for a given $\cot\theta_R$. For our present purpose 
that bound can be approximated by
\begin{equation}
\cot\theta_R \tan\theta_W \left({M_W \over M_{Z^\prime}}\right) \sim 1.
\label{appbound}
\end{equation}
Within this framework there are two potentially large sources of 
FCNC. The first one, through the coupling 
$\bar{d}_i\gamma_\mu P_R d_j Z^{\prime \mu}$ which occurs at tree level 
and which also receives large one-loop corrections (enhanced by 
$\cot\theta_R$). The processes we discuss will allow us to constrain 
the right-handed mixing angles that appear in these couplings. 
There is a second operator responsible for 
FCNC of the form $\bar{d}_i\gamma_\mu P_L d_j Z^{\prime \mu}$. This operator 
first occurs at one-loop with a finite coefficient that is enhanced 
by $\cot\theta_R$. This operator is present even when there are no FCNC 
at tree-level. Because it is enhanced by $\cot\theta_R$, 
it can contribute to a low energy FCNC process at the same level 
as the ordinary electroweak penguins mediated by the $Z$ boson even 
though $M_{Z^\prime} >> M_Z$. Both of
these operators can give rise to large contributions to
$K\rightarrow \pi \nu \bar{\nu}$ and we discuss them separately 
in what follows.

\subsection{Tree Level FCNC}

\begin{figure}[tb]
\includegraphics[width=12cm]{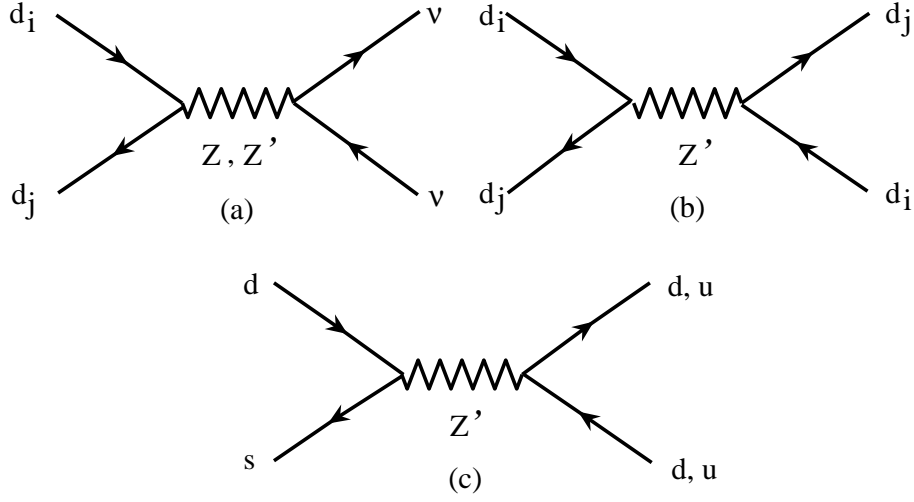}
\caption{Tree-level diagrams responsible for (a) $K \rightarrow \pi
\nu \bar{\nu}$ and $b \rightarrow (s{\rm ~or~}d) \nu \bar{\nu}$;
(b) $K-\bar{K}$ and $B-\bar{B}$ mixing; and (c) CP violation in
$K \rightarrow \pi \pi$.}
\label{fig:fdt}
\end{figure}

The models contain flavor changing neutral
currents at tree level that contribute to
$K \rightarrow \pi \nu \bar\nu$ as shown in Figure~\ref{fig:fdt}.
For models in which the third generation lepton couplings are also
enhanced, the relevant interactions are \cite{He:2003qv},
\begin{eqnarray}
{\cal L}_Z &=&  {g\over 2}\tan\theta_W (\tan\theta_R+\cot\theta_R)
(\sin\xi_Z Z_\mu + \cos\xi_Z Z^\prime_\mu) \nonumber \\
&\cdot &\left( \bar d_{Ri} \gamma^\mu V^{d*}_{Rbi} V^{d}_{Rbj} d_{Rj}
-\bar u_{Ri} \gamma^\mu V^{u*}_{Rti} V^{u}_{Rtj} u_{Rj}
+\bar \tau_R \gamma^\mu \tau_R
-\bar \nu_{R \tau} \gamma^\mu \nu_{R \tau} \right)
\label{nmcoups}
\end{eqnarray}
In this expression $g$ is the usual $SU_L(2)$ gauge coupling,
$\theta_W$ the usual electroweak angle, $\theta_R$ parametrizes
the relative strength of the right-handed interactions, $\xi_Z$
is the $Z$-$Z^\prime$ mixing angle and $V^{u,d}_{Rij}$ are the unitary matrices
that rotate the right-handed up-(down)-type quarks from the weak
eigenstate basis to the mass eigenstate basis \cite{He:2003qv}.
For $K \rightarrow \pi \nu \bar{\nu}$ Figure~\ref{fig:fdt}(a) shows
two types of FCNC:
one is mediated by ordinary $Z$ exchange and is proportional to
$\cot\theta_R \xi_Z / M_Z^2$ while the other one,
mediated by a $Z^\prime$ exchange, is proportional to $\cot^2\theta_R /
M^2_{Z^\prime}$. Eqs.~\ref{xizcon}~and~\ref{appbound} indicate that the
latter dominates so we will ignore the $Z$-$Z^\prime$ mixing from now on 
and concentrate on the simpler model with $v_L = v\cot\theta_R$ 
where $\xi_Z=0$ as mentioned above. Notice that the only enhanced 
couplings to leptons are to $\tau^\pm$ and to $\nu_\tau$. 
For this reason other rare processes such as  $K_L \rightarrow \mu^+ \mu^-$ 
are not affected in these models.

For large values of $\cot\theta_R$, the dominant tree-level  
operator responsible for $K \rightarrow \pi \nu \bar\nu$ and for 
$b \rightarrow (s{\rm ~or~}d) \nu \bar{\nu}$ is given by
\begin{equation}
{\cal L} = {g^2\tan^2\theta_W\cot^2\theta_R\over 4 M^2_{Z^\prime}}
V^{d\star}_{Rbi}V^{d}_{Rbj}\bar{d}_i\gamma_\mu
P_R d_j\  \bar{\nu}_\tau\gamma^\mu P_R \nu_\tau
+{\rm ~h.c.~}
\label{tleffl}
\end{equation}
Similarly, from Figure~\ref{fig:fdt}(b), the new $|\Delta F| = 2$ 
tree-level operators contributing to $K-\bar{K}$ and $B-\bar{B}$ mixing are 
\begin{equation}
{\cal L} = -{g^2\tan^2\theta_W\cot^2\theta_R\over 4 M^2_{Z^\prime}}
\left( V^{d\star}_{Rbi}V^{d}_{Rbj} \right)^2 \bar{d}_i\gamma_\mu
P_R d_j\  \bar{d}_i \gamma^\mu P_R d_j
+{\rm ~h.c.~}
\label{tmix}
\end{equation}
Finally, from Figure~\ref{fig:fdt}(c), the new  $|\Delta S| = 1$
operator contributing to $K \rightarrow \pi\pi$ and to $\epsilon^\prime$
is
\begin{eqnarray}
{\cal L} &=& {g^2\tan^2\theta_W\cot^2\theta_R\over 4 M^2_{Z^\prime}}
V^{d\star}_{Rbs}V^{d}_{Rbd} \bar{s}\gamma_\mu P_R d\  \nonumber \\
&&\left(V^{u\star}_{Rtu}V^{u}_{Rtu}\bar{u} \gamma^\mu P_R u
- V^{d\star}_{Rbd}V^{d}_{Rbd}\bar{d} \gamma^\mu P_R d \right)
+{\rm ~h.c.~}
\label{tk2p}
\end{eqnarray}
The phenomenological constraints that exist on flavor changing 
processes will severely restrict the non-diagonal right-handed 
mixing angles $V^d_{Rij}$ that appear above.

\subsection{Effective $\bar{d}_i\gamma_\mu P_L d_j\ Z^\prime$ operator}

\begin{figure}[tb]
\includegraphics[width=12cm]{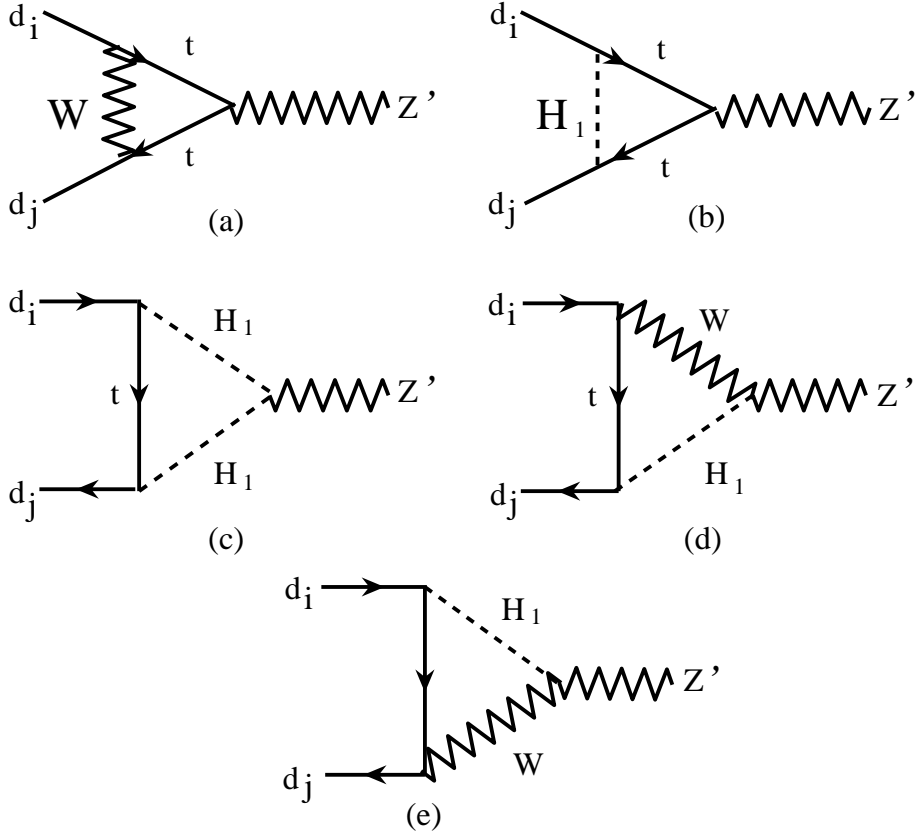}
\caption{One-loop Feynman diagrams generating
the $Z^\prime$ penguin operator of the form
$\bar{d}_i\gamma_\mu P_Ld_j Z^{\prime \mu}$ in unitary gauge.}
\label{fig:fd}
\end{figure}

A second mechanism to provide a relatively large contribution to
$K^+ \rightarrow \pi^+ \nu \bar\nu$ occurs at one-loop level and is 
present even when all the flavor changing couplings $V^d_{Rij}$
vanish. It is an effective $Z^\prime$ penguin interaction resulting from the
diagrams (in unitary gauge) shown in Figure~\ref{fig:fd}. The set
of diagrams shown corresponds to those that can be enhanced by a
large value of $\cot\theta_R$ and only arise from the intermediate
top-quark in the loop. 
The resulting low energy effective interaction is obtained by
setting the external momenta to zero in the calculation of the
loop diagrams. It can be written as
\begin{equation}
{\cal L}_{eff} = {g^3 \over 16 \pi^2}
\tan\theta_W\cot\theta_R V^\star_{ti} V_{tj}
I(\lambda_t,\lambda_H)
\bar{d}_i \gamma_\mu P_L d_j\  Z^{\prime \mu}
\label{effl}
\end{equation}
where $I(\lambda_t,\lambda_H)$ is the corresponding Inami-Lim
type function. When a third generation lepton pair is attached to 
the $Z^\prime$ a second factor $\cot\theta_R$ is introduced. The resulting 
coupling compensates for the small $M_Z/M_{Z^\prime}$ ratio and 
makes this mechanism comparable to the standard $Z$ penguin as 
follows from Eq.~\ref{appbound}. 

Additional diagrams involving a second
charged scalar that occurs in the models give rise to the operator 
$\bar{s}\gamma_\mu P_R d Z^\prime$. We do not need to consider them 
because they are simply one-loop corrections to the couplings already 
discussed in the previous section.

\begin{figure}[tb]
\includegraphics[width=10cm]{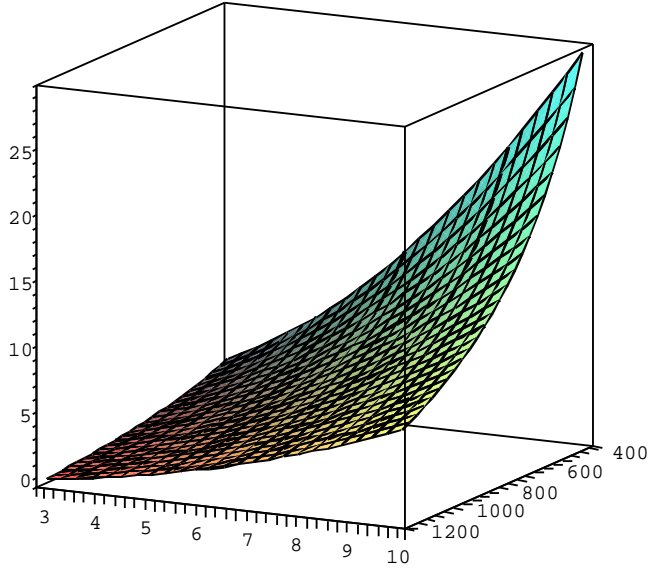}
\caption{$I(\lambda_t,\lambda_H)$ for $\cot\theta_R =3{\rm ~to~}10$ and
$M_H = 400{\rm ~to~}1200~{\rm GeV}$.}
\label{fig:formf}
\end{figure}

To compute the function $I(\lambda_t,\lambda_H)$
we work in unitary gauge and restrict ourselves to
forms of the Yukawa potential in which $b$-quark intermediate
states (and flavor changing neutral scalars) do not appear. The 
vertices needed for this calculation appear either in the appendix 
or in Ref.~\cite{He:2003qv}. Diagram~(a) in Figure~\ref{fig:fd} is 
the only one that does not involve additional parameters from the 
scalar sector of the models, it results in 
\begin{equation}
I(\lambda_t,\lambda_H)_a = {\lambda_t \over 8} \left[
{\lambda_t^2-2\lambda_t+4 \over (1-\lambda_t)^2}\log\lambda_t
-{3\over \lambda_t -1}+{1\over 2} - {1\over \hat{\epsilon}} \right].
\label{onedia}
\end{equation}
We have regularized the ultraviolet divergence in dimensional 
regularization with
\begin{equation}
{1\over \hat{\epsilon}} \equiv {2\over 4-n}-\gamma +
\log(4\pi)-\log\;\mu^2 \, .
\end{equation}
The remaining diagrams in Figure~\ref{fig:fd} cancel the divergence
and make the overall result finite but they introduce a dependence on
the Higgs parameters. For illustration we consider a 
simplified Yukawa sector detailed in the appendix in which only 
one Higgs mass appears in the result. We obtain
\begin{eqnarray}
I(\lambda_t,\lambda_H) &=& {\lambda_t \over 8} \left[
{-2\lambda_t\over (\lambda_t-\lambda_H)^2}
\left(\lambda_H\log\left({\lambda_t\over \lambda_H}\right)+\lambda_H
-\lambda_t \right)\cot^2\theta_R \right. \nonumber \\
&+&
{\lambda_H^2 (\lambda_t^2-2\lambda_t+4)
+3\lambda_t^2(2\lambda_t-2\lambda_H-1)
\over  (\lambda_t-1)^2 (\lambda_t-\lambda_H)^2}\log\lambda_t
\nonumber \\
&-& {\lambda_H(\lambda_H^2+6\lambda_t-7\lambda_H)\over
(\lambda_H-1)(\lambda_t-\lambda_H)^2}\log\lambda_H
+ \left. {2\lambda_t^2-3\lambda_H\lambda_t-5\lambda_t+6\lambda_H
\over(\lambda_t-1) (\lambda_t-\lambda_H)} \right].
\label{loopff}
\end{eqnarray}
This form factor is shown in Figure~\ref{fig:formf} for values of 
$\cot\theta_R$ between 3 and 10 and of $M_H$ between 400 and
1200~GeV. 
Within this simplified Yukawa sector the new $Z^\prime$ penguin has the 
same flavor dependence as the standard model $Z$ penguin, namely  
$V^\star_{ti}V_{tj}$. 

\section{Constraints from Mixing and CP Violation}

\subsection{$K-\bar{K}$ Mixing}

In many cases tree-level FCNC are severely constrained by
$K-\bar{K}$ mixing as in the second diagram of Figure~\ref{fig:fdt}.
In our models Eq.~\ref{tleffl} generates a new contribution to the
$K_L-K_S$ mass difference:
\begin{equation}
\Delta M = {g^2\tan^2\theta_W\cot^2\theta_R \over 4 M^2_{Z^\prime}}
{2\over 3} F_K^2 M_K B_K {\rm Re}\left(V^{d\star}_{Rbs}V^{d}_{Rbd}\right)^2.
\end{equation}
The calculation of $\Delta M$ in the kaon system is complicated by
the existence of long distance contributions.
To constrain the new interaction
we will require that its contribution to the $K_L -K_S$ mass difference
be at most equal to the largest short distance
contribution in the standard model, which is given by \cite{Buchalla:1995vs}
\begin{equation}
\Delta M \approx {G_F^2 \over 6 \pi^2} F^2_K B_K M_K M_W^2
\left(V^\star_{cs}V_{cd}\right)^2 \eta_1
{m_c^2\over M_W^2}.
\end{equation}
In this expression the QCD correction factor $\eta_1$ is equal to
$1.38$, the kaon decay constant $F_K$ and $B_K$ are as in the
notation of Ref.~\cite{Buchalla:1995vs}. For our numerical
estimates we will use the values of the CKM parameters found in
Ref.~\cite{Battaglia:2003in} in the Wolfenstein parameterization:
\begin{equation}
\lambda = 0.224,\ A= 0.839,\ \rho = 0.178,\ \eta = 0.341
\end{equation}
With all this, we obtain a constraint on the right-handed mixing angles
\begin{equation}
{\rm Re}\left(V^{d\star}_{Rbs}V^{d}_{Rbd}\right)^2
\cot^2\theta_R \tan^2\theta_W \left({M_W \over M_{Z^\prime}}\right)^2
\leq 2.4\times 10^{-8}.
\end{equation}
If we further use Eq.~\ref{appbound}, and we assume that
any new phases are small, this implies that
\begin{equation}
{\rm Re}\left(V^{d\star}_{Rbs}V^{d}_{Rbd}\right) \leq
1.5 \times 10^{-4}.
\label{kkcon}
\end{equation}

\subsection{$B-\bar{B}$ Mixing}

Similarly we consider the new contributions to mixing in the
$B$-sector from Eq.~\ref{tmix}. The calculation of $\Delta M$
in the $B_d$ system is much more precise than in the kaon system,
and the result agrees rather well with the measurement. In this
comparison, the dominant uncertainty occurs in the parameters
$F_{B_d}$ and $B_{B_d}$ that occur in the calculation.
Following Ref.~\cite{Battaglia:2003in} we write
\begin{equation}
(\Delta M)_{B_d} = 0.50 {\rm ~ps}^{-1} (1\pm 0.15)^2,
\end{equation}
which corresponds to taking $F_{B_d}\sqrt{B_{B_d}}= 230 \pm 35$~MeV,
and ignoring all other uncertainties. This is to be compared to
the world average 
$(\Delta M)_{B_d}=(0.503\pm0.006){\rm ps}^{-1}$ \cite{Battaglia:2003in}. 
In view of this
we will demand that the new physics contribution be smaller than
the theoretical uncertainty. With these numbers and Eq.~\ref{tmix}
we find,
\begin{equation}
((\Delta M)_{B_d})_{new} =
\cot^2\theta_R \tan^2\theta_W \left({M_W \over M_{Z^\prime}}\right)^2
\left|V^{d\star}_{Rbb} V^d_{Rbd}\right|^2
3.2 \times 10^{-6}~{\rm GeV} \leq 10^{-13}~{\rm GeV}.
\label{bdmix}
\end{equation}
Once again, we may combine this result with Eq.~\ref{appbound}
to find
\begin{equation}
\left|V^{d\star}_{Rbb} V^d_{Rbd}\right| \leq 1.8 \times 10^{-4}.
\label{bdmixcon}
\end{equation}

For $(\Delta M)_{B_s}$ there exists only a lower bound. We first 
consider the ratio of the new contribution to the 
standard model where the latter is obtained from the effective Hamiltonian
\cite{Buchalla:1995vs}
\begin{equation}
{\cal H}_{SM} = {G^2_f\over 4 \pi^2}M^2_W\left(V^\star_{tb}V_{ts}\right)^2
\eta_{2B} S_0(x_t)\ \bar{b}\gamma_\mu P_L s\ \bar{b} \gamma^\mu P_L s.
\end{equation}
Using $S_0(x_t) \approx 2.56$ and $\eta_{2B}= 0.55$ \cite{Buchalla:1995vs}
this results in
\begin{equation}
{(\Delta M)_{B_{s}-new}\over (\Delta M)_{B_{s}-SM}} \approx
530 \left|{V^{d\star}_{Rbb} V^d_{Rbs} \over V^\star_{tb}V_{ts}}\right|^2
\cot^2\theta_R \tan^2\theta_W \left({M_W \over M_{Z^\prime}}\right)^2. 
\label{bmixrat}
\end{equation}
We can simplify this further by using $V_{tb}\approx 1$,
$V^d_{Rbb} \approx 1$ (as discussed in Ref.~\cite{He:2003qv}), and
Eq.~\ref{appbound}. 
The eventual measurement of $B_s$ mixing will impact the modes 
that we discuss next. For this reason it will be convenient to 
define a parameter $\delta_{B_s}$ characterizing the deviation 
from the standard model prediction
$(\Delta M)_{B_s}= 17.2\ {\rm ~ps}^{-1}$ \cite{Battaglia:2003in}, 
\begin{equation}
\delta_{B_s} = \left( {(\Delta M)_{B_s} - 17.2 {\rm ~ps}^{-1} \over 
17.2 {\rm ~ps}^{-1}} \right).
\label{deldef}
\end{equation}
If we then require the new contribution to explain any deviation 
from the SM we obtain
\begin{equation}
|V^d_{Rbs}| \leq 1.8 \times 10^{-3}\ \sqrt{\delta_{B_s}} . \label{bsmixcon}
\end{equation}
Alternatively, we can use $F_{B_s}/F_{B_d}= 1.2$ to write
\begin{equation}
{(\Delta M)_{B_{s}-new}\over (\Delta M)_{B_{d}-new}} \approx
1.5 \left|{V^d_{Rbs}\over V^d_{Rbd}}\right|^2.
\end{equation}

\subsection{CP violation in $\epsilon$ and $\epsilon^\prime$}

We can also place bounds on the phases of the right-handed mixing
angles by considering the contributions of Eqs.~\ref{tmix}~and~\ref{tk2p}
to $\epsilon$ and $\epsilon^\prime$ respectively. From Eq.~\ref{tmix}
\begin{equation}
{\rm Im} M_{12} = {g^2\tan^2\theta_W \cot^2\theta_R \over 4 M^2_{Z^\prime}}
{2\over 3}F_K^2 B_K M_K {\rm Im}\left(V^{d\star}_{Rbs}V^{d}_{Rbd}\right)^2.
\end{equation}
Within the standard model the theoretical uncertainty arises predominantly
from $B_K$ and is about $30\%$.  Consequently we will
compare the new contribution to the corresponding expression in
the standard model \cite{Buchalla:1995vs} and demand that
\begin{equation}
{\epsilon_{new}\over \epsilon_{SM}} \approx 1.7 \times 10^{10}
\cot^2\theta_R \tan^2\theta_W \left({M_W \over M_{Z^\prime}}\right)^2
{\rm Re}\left(V^{d\star}_{Rbs}V^{d}_{Rbd}\right)
{\rm Im}\left(V^{d\star}_{Rbs}V^{d}_{Rbd}\right) \leq 0.3.
\end{equation}
With the aid of Eq.~\ref{appbound} this then implies that
\begin{equation}
{\rm Re}\left(V^{d\star}_{Rbs}V^{d}_{Rbd}\right)
{\rm Im}\left(V^{d\star}_{Rbs}V^{d}_{Rbd}\right) \leq 2 \times 10^{-11}.
\label{epcon}
\end{equation}

We turn now to the effect of Eq.~\ref{tk2p} on $\epsilon^\prime$.
In this case there are large theoretical uncertainties and it is best
to treat the new contributions as a correction to the result obtained
from the standard penguin operator  $\epsilon^\prime_{6}$.
We write
\begin{equation}
{\epsilon^\prime \over \epsilon} =
\left({\epsilon^\prime \over \epsilon}\right)_6 \left(1 -\Omega_{SM} -\Omega_{new}
\right).
\end{equation}
The parameter $\Omega_{new}$ contains the corrections from new physics
and we follow Ref.~\cite{He:1995na} to write Eq.~\ref{tk2p} as
\begin{equation}
{\cal H}_{new} = {g^2\over \Lambda^2}\left( \lambda_{dd} {\cal O}^{(1)}_{dd}
- \lambda_{uu} {\cal O}^{(1)}_{ud}\right).
\end{equation}
We thus identify $\Lambda = M_{Z^\prime}$ and
\begin{eqnarray}
\lambda_{dd}&=& {\tan^2\theta_W\cot^2\theta_R \over 2}
\left(V^{d\star}_{Rbs}V^{d}_{Rbd}\right)
\left|V^{d}_{Rbd}\right|^2
\nonumber \\
\lambda_{uu}&=& {\tan^2\theta_W\cot^2\theta_R \over 2}
\left(V^{d\star}_{Rbs}V^{d}_{Rbd}\right)
\left|V^{u}_{Rtu}\right|^2.
\end{eqnarray}
Using the results of Ref.~\cite{He:1995na} this leads to
\begin{equation}
\Omega_{new} = 4
\cot^2\theta_R \tan^2\theta_W \left({M_W \over M_{Z^\prime}}\right)^2
\left( 2.58 \left|V^{d}_{Rbd}\right|^2
+ 2.52 \left|V^{u}_{Rtu}\right|^2 \right)
{{\rm Im} \left(V^{d\star}_{Rbs}V^{d}_{Rbd}\right) \over
{\rm Im}\left(V^\star_{ts}V_{td}\right) }.
\end{equation}
We require this correction to be less than one and we use Eq.~\ref{appbound}
to obtain
\begin{equation}
\left(\left|V^{d}_{Rbd}\right|^2
+ \left|V^{u}_{Rtu}\right|^2 \right)
{\rm Im} \left(V^{d\star}_{Rbs}V^{d}_{Rbd}\right) \leq
1.3\times 10^{-5}.
\end{equation}

\section{Decays $K \rightarrow \pi \nu \bar{\nu}$}

To quantify the new contributions to the rate $K \rightarrow \pi \nu \bar\nu$
that occur in these models, it will be convenient to compare them to the
dominant standard model amplitude. This one arises from a top-quark
intermediate state through
the effective Hamiltonian \cite{Buchalla:1995vs}
\begin{equation}
\tilde{H}_{eff} = {G_F \over \sqrt{2}}{2\alpha\over \pi \sin^2\theta_W}
V^\star_{ts}V_{td} X(x_t)\bar{s}\gamma_\mu P_L d \
\sum_\ell \bar{\nu_\ell}\gamma^\mu P_L \nu_\ell .
\label{comp}
\end{equation}
In this expression $x_t =m_t^2/M_W^2$ and
the Inami-Lim function $X(x_t)$ is approximately equal to 1.6
\cite{Buchalla:1995vs}.

\subsection{$K^+ \rightarrow \pi^+ \bar{\nu}\nu$}

We first consider the tree-level operator of Eq.~\ref{tleffl}.
Noticing that only the vector current contributes to the
$K \rightarrow \pi$ matrix element, that the new interaction
couples only to one of the three light neutrino flavors
(the $\nu_\tau$), and that the right handed nature of the
coupling to the neutrino prevents this amplitude from
interfering with the SM amplitude, we can write
for the rate $\Gamma(K^+ \rightarrow \pi^+ \bar{\nu}\nu)$
\begin{equation}
\Gamma_{new} \approx  1130
\cot^4\theta_R \tan^4\theta_W \left({M_W \over M_{Z^\prime}}\right)^4
\left|{V^{d\star}_{Rbs}V^{d}_{Rbd}\over V^\star_{ts}V_{td}}\right|^2
\tilde{\Gamma}_{SM}.
\label{kpnnrat}
\end{equation}
We use the notation $\tilde{\Gamma}_{SM}$ to indicate that this is 
only the contribution from the top-quark intermediate state. 
If we use Eq.~\ref{appbound} as well as
the constraint from $K-\bar{K}$ mixing (Eq.~\ref{kkcon}) assuming 
that the real part dominates, 
we find that $K^+ \rightarrow \pi^+ \bar{\nu}\nu$ can be enhanced 
by two orders of magnitude with respect to the SM prediction
\begin{equation}
\Gamma_{new} \sim  200 \ \tilde{\Gamma}_{SM}.
\end{equation}
Of course this result is
in gross violation of the experimental observation, implying that
the process $K^+ \rightarrow \pi^+ \nu \bar\nu$
already places a much stronger constraint on these angles than
$K-\bar{K}$ mixing does. If we require that the
new contribution to  $K^+ \rightarrow \pi^+ \nu \bar\nu$ be at 
most as large as the standard model (which is needed to match 
the E787 and E949 central value)  we find
\begin{eqnarray}
\left|{V^{d\star}_{Rbs}V^{d}_{Rbd}\over V^\star_{ts}V_{td}}\right|^2
\cot^4\theta_R \tan^4\theta_W \left({M_W \over M_{Z^\prime}}\right)^4
&\leq& 8.9 \times 10^{-4} {\rm ~~~or,} \nonumber \\
\left|V^{d\star}_{Rbs}V^{d}_{Rbd}\right|
&\leq & 1 \times 10^{-5}.
\label{kpnncon}
\end{eqnarray}
At present these models can accommodate the 
$B(K^+\to \pi^+ \nu\bar\nu)$ obtained by E787 and E949. 
However, if we bound $|V_{Rbs}^{d*} V^d_{Rbd}|$ by a combination of 
Eqs. \ref{bdmixcon} and \ref{bsmixcon} assuming that  
$\delta_{B_s}$ is at most of order one, we find that this new contribution to
$K^+ \to \pi^+ \nu \bar \nu$ is very small. 

\begin{figure}[tb]
\includegraphics[width=15cm]{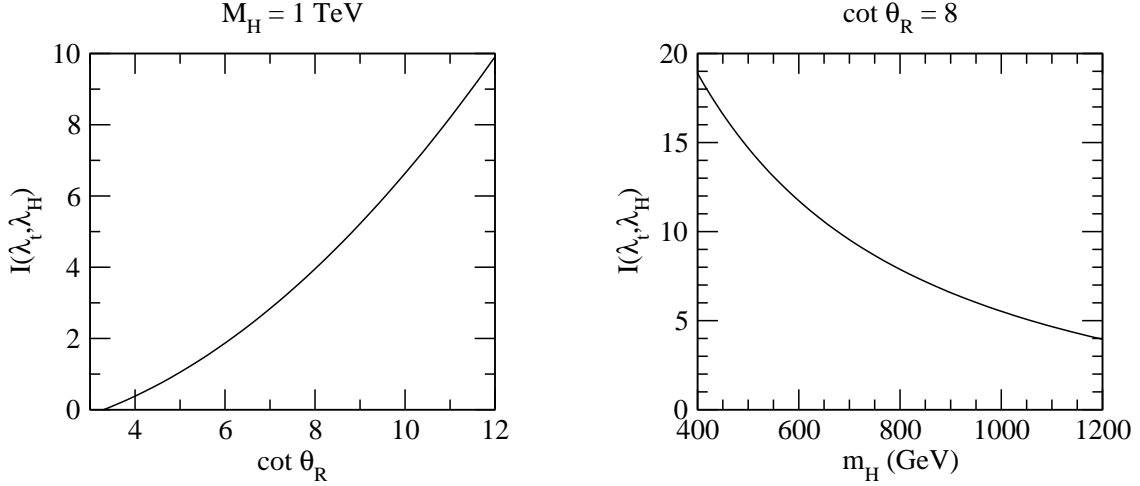}
\caption{$I(\lambda_t,\lambda_H)$ for a) $M_H=1$~TeV and
b) $\cot\theta_R = 8$.}
\label{fig:1tev}
\end{figure}

After suppressing the tree-level FCNC to a phenomenologically
acceptable level, there remains the new one-loop effective interaction
of Eq.~\ref{effl}. This effective interaction leads to a
correction to $K^+\rightarrow \pi^+ \nu \bar{\nu}$
\begin{equation}
\Gamma_{new} \sim  {1\over 12} \cot^4\theta_R \tan^4\theta_W
\left({M_W\over M_{Z^\prime}}\right)^4
\left|{I(\lambda_t,\lambda_H)\over X(x_t)}\right|^2
\tilde{\Gamma}_{SM}.
\label{loopco}
\end{equation}
Using Eq.~\ref{appbound} and requiring 
$B(K^+\to \pi^+ \nu\bar\nu)$ to be twice the SM result 
(approximately the central value of the measurement) gives 
\begin{equation}
I(\lambda_t,\lambda_H) = 5.54.
\end{equation}
From Figure~\ref{fig:formf} we see that it is possible to obtain this 
result in our model. In Figure~\ref{fig:1tev} we show two projections 
of Figure~\ref{fig:formf} for $M_H=1$~TeV and $\cot\theta_R =8$. From 
these we can read off the respective constraints that result 
from demanding that $I(\lambda_t,\lambda_H) \leq 5.54$
\begin{eqnarray}
\cot\theta_R \leq 9 & {\rm ~for~} & M_H = 1~{\rm TeV} \nonumber \\
M_H > 950~{\rm GeV} & {\rm ~for~} & \cot\theta_R = 8.
\end{eqnarray}
The form of this constraint is much more
complicated for the most general Yukawa sector. However, 
it is possible to enhance the $K^+\rightarrow \pi^+ \nu \bar{\nu}$
rate by the factor of two suggested by E787 and E949.

\subsection{$K_L \rightarrow \pi^0 \bar{\nu}\nu$}

In this case the rates are sensitive to the CP violating
component of the effective interactions. For the tree-level
operators 
\begin{equation}
\Gamma_{new} \sim  1130
\cot^4\theta_R \tan^4\theta_W \left({M_W \over M_{Z^\prime}}\right)^4
\left({ {\rm Im}(V^{d\star}_{Rbs}V^{d}_{Rbd})\over
{\rm Im}(V^\star_{ts}V_{td})}\right)^2
\tilde{\Gamma}_{SM}.
\end{equation}
Using the current experimental bound
$B(K_L \rightarrow \pi^0 \bar{\nu}\nu) < 5.9 \times 10^{-7}$
\cite{Alavi-Harati:1999hd}, combined with the standard model 
prediction (central value) 
$B(K_L \rightarrow \pi^0 \bar{\nu}\nu) = 2.4 \times 10^{-11}$
\cite{Battaglia:2003in} and Eq.~\ref{appbound} results in
\begin{equation}
\left|{\rm Im}\left(V^{d\star}_{Rbs}V^{d}_{Rbd}\right)\right| \leq
6.3 \times 10^{-4}.
\end{equation}
Alternatively, if we use Eq.~\ref{kpnncon}, assuming that the 
imaginary part saturates the $K^+ \rightarrow \pi^+ \bar{\nu}\nu$ 
rate, we find
\begin{equation}
B(K_L \rightarrow \pi^0 \bar{\nu}\nu) < 1.4 \times 10^{-10}.
\end{equation}

As mentioned before, the one-loop operator (Eq.~\ref{loopff})  
has the same phase as the standard model effective
Hamiltonian, Eq.~\ref{comp} and therefore its contribution 
to $K_L \rightarrow \pi^0 \bar{\nu}\nu$ is 
rigorously limited by its 
contribution to $K^+ \rightarrow \pi^+ \bar{\nu}\nu$. 
If this is the only new physics present
\begin{equation}
\Gamma(K_L \rightarrow \pi^0 \bar{\nu}\nu) = 
\Gamma(K_L \rightarrow \pi^0 \bar{\nu}\nu)_{SM} \left(
{\Gamma(K^+\rightarrow \pi^+ \bar{\nu}\nu) \over 
\Gamma(K^+\rightarrow \pi^+ \bar{\nu}\nu)}_{SM}\right).
\end{equation}

\section{Rare B decay modes}

The $K$ decay modes discussed in the previous
section have obvious extensions to the $B$ sector. In addition to that
there are additional modes involving the third generation charged lepton
that are also enhanced in the models we discuss. The inclusive
modes of the form $B \rightarrow X \tau^+ \tau^-$ receive contributions
from an intermediate photon and are less sensitive to the $Z^\prime$ operators 
so we will not discuss them. Additional $B$ decay modes have been 
considered within the context of FCNC and $Z^\prime$ \cite{Barger:2003hg}.  
These additional modes involve first and second generation fermions 
coupling to the $Z^\prime$ and for this reason they do not constrain 
our model significantly.

\subsection{$B \rightarrow X_{d,s} \nu \bar{\nu}$}

For the modes $B \rightarrow X_{d,s} \nu \bar{\nu}$ the analogue to 
Eq.~\ref{kpnnrat} for the tree-level FCNC is
\begin{equation}
{\Gamma_{new}\over \Gamma_{SM}} \approx 1130
\cot^4\theta_R \tan^4\theta_W \left({M_W \over M_{Z^\prime}}\right)^4
\left|{V^{\star d}_{Rbb}V^d_{Rbi} \over V^\star_{tb}V_{ti}}\right|^2.
\label{bxnnrat}
\end{equation}
For $B \rightarrow X_d \nu \bar{\nu}$ we can use the constraint
Eq.~\ref{bdmixcon} together with Eq.~\ref{appbound} to see that
the additional new contribution does not change the standard model
prediction by more than $50\%$. Using the central value 
$B_{SM}(B \rightarrow X_d \nu \bar{\nu}) = 1.6 \times 10^{-6}$ 
\cite{Buchalla:1995vs} this implies
\begin{equation}
B(B \rightarrow X_d \nu \bar{\nu}) \leq 2.4 \times 10^{-6}.
\end{equation}
Similarly we can use 
$B_{SM}(B \rightarrow X_s \nu \bar{\nu}) = 4 \times 10^{-5}$ 
\cite{Buchalla:1995vs}, along with Eqs.~\ref{bmixrat},~\ref{bxnnrat} 
to obtain in terms of Eq.~\ref{deldef}
\begin{equation}
B(B \rightarrow X_s \nu \bar{\nu}) \leq 4 \times 10^{-5}
(1+2\ \delta_{B_s}).
\end{equation}
At one-loop the form of Eqs.\ref{comp}~and~\ref{loopff} implies
that these modes receive identical constraints to Eq~\ref{loopco}. 

\subsection{$B_i \rightarrow \tau^+\tau^-$}

Finally the effective Hamiltonian responsible for
$B_i \rightarrow \tau^+\tau^-$
within the standard model is \cite{Buchalla:1995vs}
\begin{equation}
{\cal H} = -{G_F \over \sqrt{2}}{2\alpha \over \pi \sin^2\theta_W}
V^\star_{tb}V_{ti}\ Y(x_t)\ \bar{b}\gamma_\mu P_L d_i\ \bar\tau
\gamma^\mu P_L \tau
\label{bllsm}
\end{equation}
where the Inami-Lim function $Y(x_t)\approx  1.06$ \cite{Buchalla:1995vs}.
Using the central values for hadronic parameters from
Ref.~\cite{Battaglia:2003in} this leads to standard model branching ratios
$B(B_s\rightarrow \tau^+\tau^-) = 1.1 \times 10^{-6}$ and
$B(B_d\rightarrow \tau^+\tau^-) = 3.3 \times 10^{-8}$. The contribution
of the tree-level FCNC to the rates, normalized by the 
standard model rate is
\begin{equation}
{\Gamma_{new}\over \Gamma_{SM}} \approx 7730
\cot^4\theta_R \tan^4\theta_W \left({M_W \over M_{Z^\prime}}\right)^4
\left|{V^{\star d}_{Rbb}V^d_{Rbi} \over V^\star_{tb}V_{ti}}\right|^2.
\end{equation}
This is somewhat larger than the corresponding ratio for
$b\rightarrow d_i \nu \bar{\nu}$ because it does not have the factor
of $1/3$ that indicates that only the $\nu_\tau$ couples strongly to
the new interactions and also because $Y(x_t)$ is slightly smaller than
$X(x_t)$. With the constraints Eqs.~\ref{appbound}~and~\ref{bdmixcon}
we find that
\begin{equation}
B(B_d \rightarrow \tau^+ \tau^-) \leq 1.5 \times 10^{-7}.
\end{equation}
For the $B_s$ we resort again to Eq.~\ref{deldef} to write
\begin{equation}
B(B_s \rightarrow \tau^+ \tau^-) = 1.1 \times 10^{-6} 
\left( 1 + 15\ \delta_{B_s} \right) .
\end{equation}

The contribution to these rates from the new penguin operator,  
relative to their standard model value, 
is about a factor of 7 larger that its contribution to the 
$K^+ \rightarrow \pi^+ \nu \bar\nu$ mode. Again, this factor 
arises from $Y(x_t)$ being somewhat smaller than $X(x_t)$ and 
because the new operator only couples to $\nu_\tau$.

\section{Summary and Conclusions}

We have examined the effects of a new non-universal
right-handed $Z^\prime$ boson on several rare flavor changing
processes: $K^+\to \pi^+ \nu\bar\nu$, $K_L\to \pi^0 \nu\bar\nu$,
$K-\bar{K}$ and $B-\bar{B}$ mixing, $B \rightarrow X_{d,s} \nu
\bar{\nu}$ and $B\rightarrow \tau^+ \tau^-$. There are two 
mechanisms by which these models can enhance these modes with 
respect to the standard model.

The first mechanism arises from tree-level FCNC present in the 
couplings of the non-universal right-handed $Z^\prime$ boson. 
The new $Z^\prime$ boson can only have large couplings to third 
generation leptons and its couplings to the $\tau$ lepton are 
constrained by LEP. In the quark sector, the flavor conserving 
couplings of the $Z^\prime$ are large only for the $b$ and $t$ 
quarks. However the flavor changing couplings to the first 
two generation quarks could be large if the right-handed 
mixing angles are large. In this paper we obtained stringent bounds
on these mixing parameters using known experimental data on $\Delta
M_K$, $\epsilon_{K}$, $B_d$~mixing,  $\epsilon'/\epsilon_K$ 
and $K^+\to \pi^+ \nu\bar\nu$ which we summarize in 
Table~\ref{t:constraints}.

\begin{table}[tbh]
\centering
\caption[]{Summary of constraints for the right-handed 
mixing angles.}
\begin{tabular}{|l|l|c|} \hline
Process & Constraint &  \\ \hline
$(\Delta M)_K$ & ${\rm Re}\left(V^{d\star}_{Rbs}V^d_{Rbd}\right)^2
< 2.4 \times 10^{-8}$
& I  \\
$(\Delta M)_{B_d}$ & $\left|V^{d\star}_{Rbb}V^d_{Rbd}\right| 
< 1.8 \times 10^{-4}$& II  \\
$(\Delta M)_{B_s}$ & $\left|V^{d\star}_{Rbb}V^d_{Rbs}\right| 
< 1.8 \times 10^{-3} \sqrt{\delta_{B_s}}$& $\delta_{B_s}$  \\
$\epsilon $ & ${\rm Re}\left(V^{d\star}_{Rbs}V^d_{Rbd}\right)
{\rm Im}\left(V^{d\star}_{Rbs}V^d_{Rbd}\right) < 2 \times 10^{-11}$& III \\
$\epsilon^\prime$ & $\left(\left|V^{d}_{Rbd}\right|^2
+ \left|V^{u}_{Rtu}\right|^2 \right)
{\rm Im} \left(V^{d\star}_{Rbs}V^{d}_{Rbd}\right) \leq
1.3\times 10^{-5}$ & IV \\
$B(K^+ \rightarrow \pi^+ \nu \bar{\nu})$ & 
$\left|V^{d\star}_{Rbs}V^d_{Rbd}\right| < 1.0 \times 10^{-5}$ & V \\ \hline
\end{tabular}
\label{t:constraints}
\end{table}

At present, the models can reproduce the central value of E787 and E949 
for the rate for $K^+ \rightarrow \pi^+ \nu \bar{\nu}$ with the 
tree-level FCNC. However this can easily change if $B_s$ mixing is 
measured and does not deviate much from the standard model. With 
all the constraints on Table~\ref{t:constraints} we can predict 
additional modes which we summarize in Table~\ref{t:predictions}.

\begin{table}[tbh]
\centering
\caption[]{Summary of Predictions.}
\begin{tabular}{|l|l|c|} \hline
Process & Prediction  & From \\ \hline
$B(K_L \rightarrow \pi^0 \nu \bar{\nu})$ & $<1.4\times 10^{-10}$ & V \\
$B(B \rightarrow X_d \nu \bar{\nu})$ & $< 2.5\times 10^{-6}$ & II \\
$B(B \rightarrow X_s \nu \bar{\nu})$ & $<4\times 10^{-5}(1+2\ \delta_{B_s})$
& $\delta_{B_s}$ \\
$B(B_d \rightarrow \tau^+ \tau^-)$ & $<1.8\times 10^{-7}$ & II \\
$B(B_s \rightarrow \tau^+ \tau^-)$ & $<1.1\times 10^{-6} (1 +15\ 
\delta_{B_s})$ 
& $\delta_{B_s}$ \\ \hline
\end{tabular}
\label{t:predictions}
\end{table}

The second mechanism to enhance the flavor changing processes occurs 
at one-loop in the form of a $\bar{d}_i \gamma_\mu P_L d_j\ Z^{\prime \mu}$ 
penguin. The models contain too many free parameters to specify 
completely the coefficient of this operator. We worked with a 
simplified model in which the coefficient can be predicted in terms 
of one Higgs mass and $\cot\theta_R$. In this 
case we found that the process $K^+\to \pi^+\nu\bar \nu$ can be 
enhanced to meet the central value of E787 and E949.  Assuming that 
this central value is correct, we can use $K^+\to \pi^+\nu\bar \nu$ 
to predict other processes. We find that $K_L \rightarrow \pi^0 \nu 
\bar\nu$ and $B \rightarrow X_{d,s} \nu \bar\nu$ are enhanced with 
respect to the standard model by exactly the same factor as 
$K^+\to \pi^+\nu\bar \nu$ with this mechanism. On the other hand 
the modes $B_{d,s} \rightarrow \tau^+ \tau^-$ are enhanced 
approximately seven times as much.

At present the models allow large deviations from the standard 
model in $B_s-\bar B_s$ mixing and in rare $B$ decays such as 
$B_{d,s} \to \tau^+\tau^-$. They can also reach the central value 
of E787 and E949 for $K^+\to \pi^+\nu\bar \nu$ both with the tree-level 
FCNC and with the new $Z^\prime$ penguin operator. A measurement of 
$B_s-\bar B_s$ mixing will most likely limit the tree-level 
FCNC contribution to  $K^+\to \pi^+\nu\bar \nu$ to a very small correction. 
The $Z^\prime$ penguin operator, however, will not be constrained by 
this measurement and can still enhance $K^+\to \pi^+\nu\bar \nu$ 
significantly.

\noindent {\bf Acknowledgments}$\,$ The work of X.G.H. was
supported in part by the National Science Council under NSC grants. 
The work of G.V. was supported in part by DOE under contract
number DE-FG02-01ER41155. G.V. thanks the School of Physics at 
UNSW for their hospitality while this work was completed.

\appendix

\section{Yukawa sector and Gauge-Boson-Scalar couplings}

The most general Yukawa potential for
these models is given by
\begin{eqnarray}
{\cal L}_Y  &=& \bar Q^i_L\lambda^u_{ij} H_L U^j_R  + \bar
Q^i_L\lambda^d_{ij} \tilde H_LD^j_R \nonumber\\
&+& \bar Q^i_L \lambda_i \phi Q^3_R + \bar Q^i_L \tilde
\lambda_i \tilde \phi Q^3_R,
\end{eqnarray}
where $i=1,2,3$, $j=1,2$, $\tilde H_L = i\tau_2 H^*_L$, $\tilde \phi = i\tau_2
\phi^* i\tau_2$ and we use the notation of Ref.~\cite{He:2003qv}
for all fields.
This Lagrangian contains too many free parameters so we restrict
ourselves to a simpler case. We pick $\lambda_{1,2}=0$ and
$\tilde\lambda_{1,2}=0$. 
These choices imply that $m_t = -\lambda_3 v$, $m_b = \tilde\lambda_3 v$
and the necessary fermion-scalar
couplings become
\begin{equation}
{\cal L}_Y =  -{m_t \over v}
\cos\theta_R \left( V^\star_{tb}\bar{b}_L t_R H_1^- +
V^\star_{ts}\bar{s}_L t_R H_1^- + V^\star_{td}\bar{d}_L t_R H_1^-
\right).
\label{yukphy}
\end{equation}
The vertices of the form $Z HH$ not already listed in Ref.~\cite{He:2003qv}
are obtained from the Lagrangian
\begin{eqnarray}
{\cal L} &=&
-  {i g_L \tan\theta_W \over 2\cos\theta_R\sin\theta_R } \left(
{v_R^2\cos^2\theta_R+2v^2\cos^2\theta_R-v^2\over v_R^2+v^2}
Z^{\prime \mu}\left(H_2^-\partial_\mu H_2^+ - H_2^+\partial_\mu H_2^-\right)
\right. \nonumber \\
&+& \left.
(2\cos^2\theta_R-1)
Z^{\prime \mu}\left(H_1^-\partial_\mu H_1^+ - H_1^+\partial_\mu H_1^-\right)
\right).
\end{eqnarray}
Finally, the vertices of the form $Z^\prime WH$ are obtained from the
Lagrangian
\begin{equation}
{\cal L} = -{g_L^2\over \sqrt{2}} \left(
{vv_R\over \sqrt{v_R^2+v^2}}{\tan^2\theta_W \over \cos\theta_R}
Z^{\prime \mu} W_{R\mu}^\pm H^\mp_2 + {v \tan\theta_W \over \sin\theta_R}
Z^{\prime \mu} W_{L\mu}^\pm H^\mp_1\right).
\end{equation}


\begin{thebibliography}{99}

\bibitem{Adler:2001xv}
S.~Adler {\it et al.}  [E787 Collaboration],
Phys.\ Rev.\ Lett.\  {\bf 88}, 041803 (2002)
[arXiv:hep-ex/0111091].

\bibitem{Artamonov:2004hr}
A.~V.~Artamonov  [E949 Collaboration],
arXiv:hep-ex/0403036.

\bibitem{Buchalla:1995vs}
G.~Buchalla, A.~J.~Buras and M.~E.~Lautenbacher,
Rev.\ Mod.\ Phys.\  {\bf 68}, 1125 (1996) [arXiv:hep-ph/9512380].

\bibitem{Battaglia:2003in}
M.~Battaglia {\it et al.},
arXiv:hep-ph/0304132.

\bibitem{smz}
A.~J.~Buras, R.~Fleischer, S.~Recksiegel and F.~Schwab,
arXiv:hep-ph/0402112.

\bibitem{susy}
A.~J.~Buras, A.~Romanino and L.~Silvestrini,
Nucl.\ Phys.\ B {\bf 520}, 3 (1998) [arXiv:hep-ph/9712398];
G.~Colangelo and G.~Isidori,
JHEP {\bf 9809}, 009 (1998) [arXiv:hep-ph/9808487]; A.~J.~Buras,
G.~Colangelo, G.~Isidori, A.~Romanino and L.~Silvestrini,
Nucl.\ Phys.\ B {\bf 566}, 3 (2000) [arXiv:hep-ph/9908371];
C.~H.~Chen,
J.\ Phys.\ G {\bf 28}, L33 (2002) [arXiv:hep-ph/0202188]; Y.~Nir
and G.~Raz,
Phys.\ Rev.\ D {\bf 66}, 035007 (2002) [arXiv:hep-ph/0206064];
S.~Baek, J.~H.~Jang, P.~Ko and J.~H.~Park,
Nucl.\ Phys.\ B {\bf 609}, 442 (2001)
[arXiv:hep-ph/0105028].

\bibitem{z}
J.A. Aguilar-Saavedra, Phys. Rev. {\bf D67}, 035003(2003)
[arXiv:hep-ph/0210112].

\bibitem{other}
W.~F.~Chang and J.~N.~Ng,
JHEP {\bf 0212}, 077 (2002) [arXiv:hep-ph/0210414]; G.~D'Ambrosio,
G.~F.~Giudice, G.~Isidori and A.~Strumia,
Nucl.\ Phys.\ B {\bf 645}, 155 (2002) [arXiv:hep-ph/0207036];
G.~Burdman,
Phys.\ Rev.\ D {\bf 66}, 076003 (2002) [arXiv:hep-ph/0205329];
D.~Hawkins and D.~Silverman,
Phys.\ Rev.\ D {\bf 66}, 016008 (2002) [arXiv:hep-ph/0205011];
T.~Yanir,
JHEP {\bf 0206}, 044 (2002) [arXiv:hep-ph/0205073];A.~J.~Buras,
M.~Spranger and A.~Weiler,
Nucl.\ Phys.\ B {\bf 660}, 225 (2003) [arXiv:hep-ph/0212143].

\bibitem{Buchalla:1995dp}
G.~Buchalla, G.~Burdman, C.~T.~Hill and D.~Kominis,
Phys.\ Rev.\ D {\bf 53}, 5185 (1996) [arXiv:hep-ph/9510376].

\bibitem{He:2002kn}
X.~G.~He and G.~Valencia,
Phys.\ Rev.\ D {\bf 66}, 013004 (2002)
[Erratum-ibid.\ D {\bf 66}, 079901 (2002)].

\bibitem{He:2003qv}
X.~G.~He and G.~Valencia,
Phys.\ Rev.\ D {\bf 68}, 033011 (2003)
[arXiv:hep-ph/0304215].

\bibitem{Chanowitz:2001bv}
M.~S.~Chanowitz,
Phys.\ Rev.\ Lett.\  {\bf 87}, 231802 (2001)
[arXiv:hep-ph/0104024].

\bibitem{Chanowitz:1999jj}
M.~S.~Chanowitz,
arXiv:hep-ph/9905478.

\bibitem{Abbaneo:2001ix}
D.~Abbaneo {\it et al.}  [ALEPH Collaboration],
arXiv:hep-ex/0112021.

\bibitem{He:1995na}
X.~G.~He and G.~Valencia,
Phys.\ Rev.\ D {\bf 52}, 5257 (1995) [arXiv:hep-ph/9508411].

\bibitem{Alavi-Harati:1999hd}
A.~Alavi-Harati {\it et al.}  [The E799-II/KTeV Collaboration],
Phys.\ Rev.\ D {\bf 61}, 072006 (2000) [arXiv:hep-ex/9907014].

\bibitem{Barger:2003hg}
V.~Barger, C.~W.~Chiang, P.~Langacker and H.~S.~Lee,
Phys.\ Lett.\ B {\bf 580}, 186 (2004)
[arXiv:hep-ph/0310073].


\end{thebibliography}
\end{document}